\documentclass{article}
\usepackage{spconf,amsmath,graphicx}
\usepackage{hyperref}

\usepackage{booktabs}      
\usepackage{multirow}      
\usepackage{graphicx}      
\usepackage{array}         

\usepackage{booktabs} 
\usepackage{graphicx} 
\usepackage{amssymb}

\title{OSUM-Pangu: An Open-Source Multidimension Speech Understanding Foundation Model Built upon OpenPangu on Ascend NPUs}

\name{Yujie Liao$^*$, Xuelong Geng$^*$, Hongfei Xue, Shuiyuan Wang, Lei Xie$^\dag$\thanks{$^*$Equal contribution. $^\dag$Corresponding author.}}

\address{
Audio, Speech and Language Processing Group \href{http://www.npu-aslp.org}{(ASLP@NPU)},\\
Northwestern Polytechnical University, Xi'an, China \\
\textbf{Project page:} \url{https://github.com/ASLP-lab/OSUM-Pangu}
}

\begin{document}
\ninept
\maketitle

\begin{abstract}
Recent advancements in Speech Large Language Models have significantly enhanced multi-dimensional speech understanding. However, the majority of high-performance frameworks are predominantly optimized for GPU centric ecosystems and proprietary backbones, creating a significant gap for deployment on non-CUDA computing infrastructures. In this paper, we present \textbf{OSUM-Pangu}, a fully open-source speech understanding foundation model developed on a completely non-CUDA software and hardware stack. By integrating an audio encoder with the openPangu-7B LLM backbone, we successfully implement the entire training and inference pipeline on the Ascend NPU platform. To facilitate efficient task alignment under non-CUDA resource constraints, we adopt a practical training process that sequentially bridges speech perception and user intent recognition. Experimental results demonstrate that OSUM-Pangu achieves task accuracy comparable to mainstream GPU-based models while maintaining robust natural language interaction capabilities. Our work provides a reproducible, non-CUDA baseline for the open-source speech community, promoting the independent evolution of multimodal intelligence.
\end{abstract}

\section{Introduction}
\label{sec:intro}

The rapid advancement of Large Language Models (LLMs) has significantly reshaped the landscape of Artificial Intelligence, demonstrating strong capabilities in complex reasoning, zero-shot generalization, and open-domain instruction following \cite{llm:llama3,llm:gpt4,pangu}. Building upon these advances, integrating pre-trained acoustic encoders with LLMs has emerged as a dominant paradigm for speech understanding \cite{salmonn,um:qwen-audio}. Early studies primarily focused on bridging the modality gap between acoustic signals and textual representations, enabling reliable automatic speech recognition and basic semantic understanding \cite{simple-asr-llm,asr_my}. More recently, speech systems \cite{audio_flamingo3,um:qwen2audio} have evolved toward unified multimodal frameworks that support broader perceptual capabilities beyond transcription, including emotion recognition, event detection, and gender classification.

Recent open-source models and LLMs such as Qwen2-Audio \cite{um:qwen2audio}, SALMONN \cite{salmonn}, and OSUM \cite{osum} extend this paradigm by integrating multiple speech related tasks within a single LLM-based architecture. Proprietary systems such as Gemini \cite{team2023gemini} and GPT-4o \cite{hurst2024gpt} further demonstrate strong multimodal interaction capabilities by supporting flexible instruction driven speech processing. These advances highlight a growing trend toward holistic speech intelligence, where models are expected to interpret diverse acoustic signals and execute complex tasks under natural language instructions.

Despite these advances, several challenges remain. Most speech understanding systems are developed primarily for the NVIDIA CUDA ecosystem, limiting their portability across alternative hardware platforms. To address this gap, the OSUM framework \cite{osum} establishes a robust open-source baseline by supporting multiple speech tasks efficiently and has also been adapted to the NPU platform. However, its backbone language model, Qwen2 \cite{yang2024qwen2technicalreport}, still relies on CUDA and lacks native optimization for NPU. 
In addition, OSUM \cite{osum} depends on rigid task specific prompts or fixed task labels. In practice, user intent is often expressed in free form language rather than predefined formats, which limits its ability to interpret natural language instructions.

To address these challenges, we propose \textbf{OSUM-Pangu}. We adopt OSUM \cite{osum} as the framework due to its strong speech perception capability and mature adaptation to NPU infrastructures. As the backbone, openPangu-7B \cite{pangu} is a large language model pre-trained on the Ascend NPU architecture, providing natural compatibility and optimized performance within the NPU ecosystem. By integrating OSUM with openPangu-7B, we construct a non-CUDA software hardware stack that enables the unified speech LLM paradigm on Ascend NPUs while maintaining an end-to-end multimodal training workflow comparable to CUDA centric ecosystems. Furthermore, we adopt an intent aware training strategy that improves the model’s ability to interpret natural language instructions without relying on rigid prompts or large multimodal datasets. Experiments show that OSUM-Pangu achieves competitive performance across multiple speech understanding tasks while maintaining a high instruction following rate of 90.2\%. We release the model code and weights to provide a reproducible baseline for speech understanding research in non-CUDA environments.

\section{Related Work}
\label{sec:related_work}

\textbf{Speech Understanding Models.}
The integration of speech processing with Large Language Models has evolved from cascaded pipelines to unified multimodal architectures. 
Subsequent research shifted toward Speech Large Language Models with stronger perception and reasoning capabilities. Representative works include LTU and LTU-AS \cite{gong2023ltu, gong2023ltu-as}, which introduced the “listen–think–understand’’ paradigm, Pengi \cite{deshmukh2023pengi}, a unified model supporting both closed-set and open-set audio tasks, and GAMA \cite{ghosh2024gama}, which enhanced reasoning across speech, sound, and music. Later systems such as SALMONN \cite{salmonn}, Qwen-Audio and Qwen2-Audio \cite{um:qwen-audio, um:qwen2audio}, Phi-4-mini \cite{abouelenin2025phi}, and the Audio Flamingo series \cite{kong2024audioflamingo, kong2025audioflamingo2, audio_flamingo3} further improved instruction following, long-audio modeling, and multimodal interaction. To further improve instruction-following capability, recent studies \cite{xue2025enhancing} have explored techniques such as Chain-of-Thought reasoning., while closed-source models like Gemini \cite{team2023gemini} and GPT-4o \cite{hurst2024gpt} set new performance benchmarks. Among open-source frameworks, OSUM \cite{osum} and OSUM-EChat \cite{osum-echat} provide competitive baselines for unified speech understanding with native support for acoustic modeling and cross-modal alignment on Ascend NPUs. However, the backbone language model used by OSUM still relies on the CUDA ecosystem and lacks native architecture optimization for the NPU platform. Our work builds upon this foundation by integrating OSUM with the openPangu backbone to advance speech understanding on non-CUDA heterogeneous platforms.

\textbf{NPU Infrastructure and Ecosystem.}
With the growing demand for diversified AI computing architectures, the Ascend NPU platform has emerged as a scalable infrastructure for large-scale model training and deployment. Prior studies have demonstrated the feasibility of running mainstream frameworks such as MindSpore and PyTorch on Ascend NPUs \cite{zhu2023performance}, while distributed optimizations enable efficient scaling across multi-NPU clusters \cite{wang2023analysis}. Within this ecosystem, the openPangu family \cite{pangu, yin2025pangu} represents a major milestone in developing LLMs optimized for Ascend hardware, with openPangu Ultra \cite{yin2025pangu} demonstrating extreme-scale training using thousands of NPUs. However, despite these advances in text-based models, multimodal research particularly speech understanding remains limited on the Ascend platform, as most Audio-LLMs rely on CUDA-based infrastructures. To address this gap, we integrate the NPU compatible OSUM framework with the openPangu backbone, resulting in OSUM-Pangu, a open-source multidimension speech foundation model designed for non-CUDA environments.

\begin{figure}[tb] 
    \centering
    \makebox[\linewidth][c]{
        \includegraphics[width=\linewidth, trim=0 0 0 0, clip]{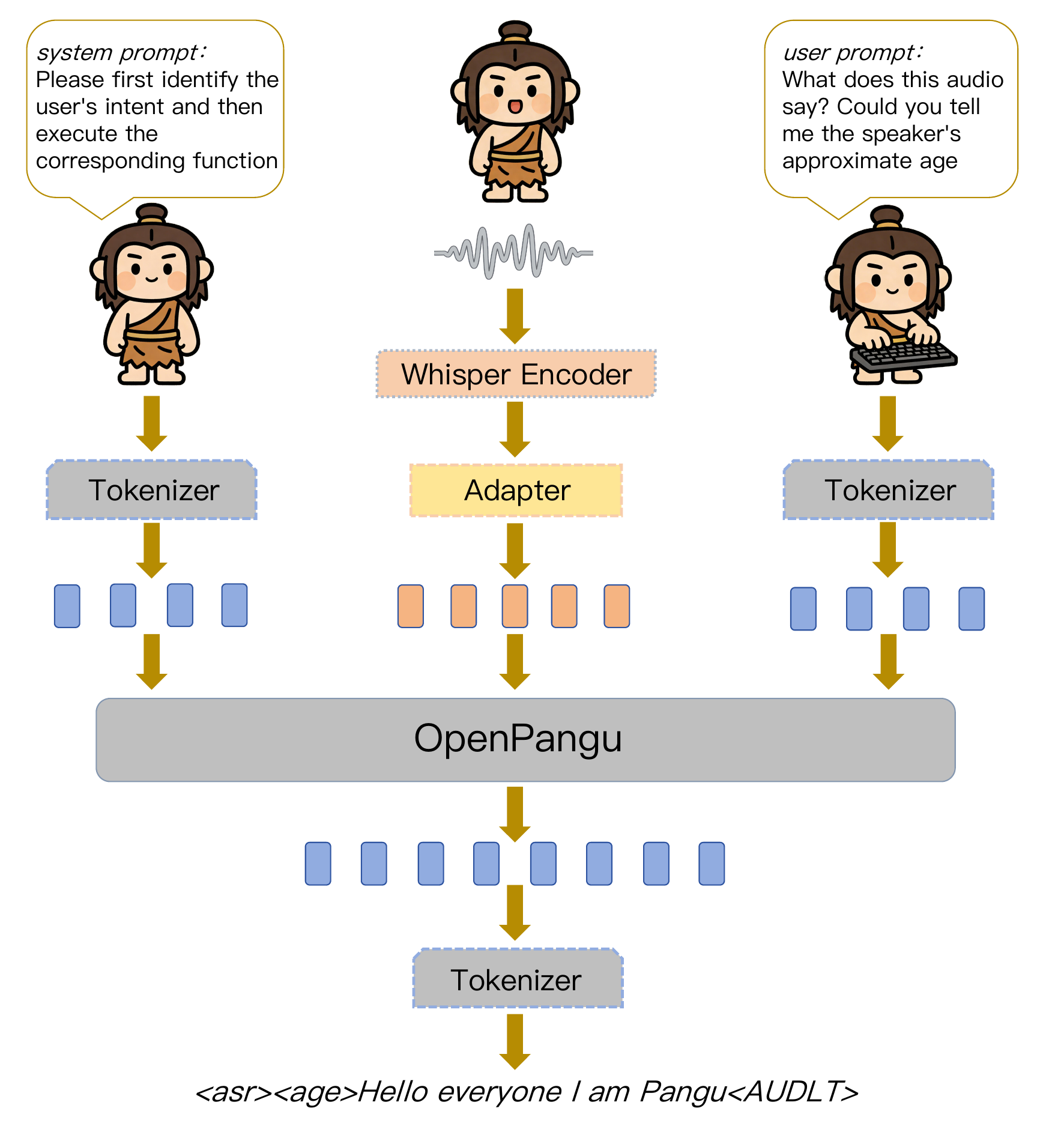}
    }
    \caption{\textbf{Illustration of the OSUM-Pangu model inference workflow.} 
    The architecture processes text instructions (left/right) and speech signals (middle) through a Transformer-based adapter and openPangu backbone, producing structured outputs with task-specific tags.}
    \label{fig:model_inference}
\end{figure}

\section{Method}
\label{sec:method}

To achieve high  performance speech understanding in non-CUDA software hardware environments, we propose the OSUM Pangu speech understanding framework based on the NPU platform adapted speech understanding framework OSUM.The framework leverages the synergy between the openPangu language backbone and the Ascend NPU hardware platform. To ensure efficient convergence and task alignment under this system, we adopt a multi-stage training pipeline that integrates acoustic perception with linguistic reasoning.

\subsection{Model Architecture}

As illustrated in Fig.~\ref{fig:model_inference}, OSUM-Pangu adopts a modular architecture designed for seamless deployment on non-CUDA infrastructure. It consists of three primary components: an acoustic encoder, a trainable modality adapter, and a LLM backbone.

\textbf{Acoustic Encoder:}
To achieve comprehensive speech understanding, the system requires a frontend capable of extracting both semantic content and paralinguistic attributes.We employ the pre-trained Whisper-medium encoder \cite{asr:whisper} as the acoustic foundation. Given an input audio signal $X$, it is first transformed into an 80-channel log-Mel spectrogram and then encoded into a sequence of acoustic embeddings:
\begin{equation}
H_a = \text{Enc}(X), \quad H_a \in \mathbb{R}^{T \times D_a}
\end{equation}
where $T$ denotes the temporal length and $D_a$ the hidden dimension. To ensure training stability and maximize the computational efficiency of the NPU hardware, the encoder parameters remain frozen during the initial alignment stages.

\textbf{Modality Adapter:} 
To address modality gap challenge in multimodal interaction, the modality adapter is designed to perform both feature alignment and temporal compression. Specifically, two 2D convolutional layers are used to perform $4\times$ temporal downsampling on $H_a$, reducing the sequence length to $L = T/4$. This is followed by $N$ Transformer layers to model long-range acoustic dependencies. Finally, a linear projection maps the features to the LLM embedding dimension $D_{llm}$, producing speech tokens $Z_a$:
\begin{equation}
Z_a = \text{Linear}(\text{Transformer}(\text{CNN}_{2D}(H_a)))
\end{equation}
This compression allows the LLM to process long-duration audio efficiently within its limited context window.

\textbf{LLM Backbone:} 
 To support deployment in non-CUDA software hardware environments, compared with openPangu-Ultra, openPangu-Embedded-7B-V1.1 offers lower memory footprint and computational overhead, enabling efficient integration with the acoustic encoder of the OSUM framework. Thus, we employ openPangu-Embedded-7B-V1.1\footnote{Model download link:
\href{https://huggingface.co/FreedomIntelligence/openPangu-Embedded-7B-V1.1}{openPangu-Embedded-7B-V1.1}} \cite{pangu} as the backbone language model.  During inference, the LLM processes a hybrid sequence where speech tokens $Z_a$ are inserted into natural language instruction tokens $T_{inst}$. Following the arrangement in Fig.~\ref{fig:model_inference}, the final input is:
\begin{equation}
I_{final} = [T_{inst}^{left}; Z_a; T_{inst}^{right}]
\end{equation}
where $T_{inst}^{left}$ and $T_{inst}^{right}$ denote the prefix and suffix instruction segments surrounding the speech tokens. The LLM performs auto-regressive generation to produce structured responses. During training, the model is encouraged to prepend task-specific identifiers, such as \texttt{<asr>} or \texttt{<age>} to its output. These identifiers enable the model to perform implicit task routing and structured information extraction based on the interpreted user intent.

\begin{figure}[t]
    \centering
    \makebox[\linewidth][c]{
        \includegraphics[width=\linewidth, trim=70 0 70 0, clip]{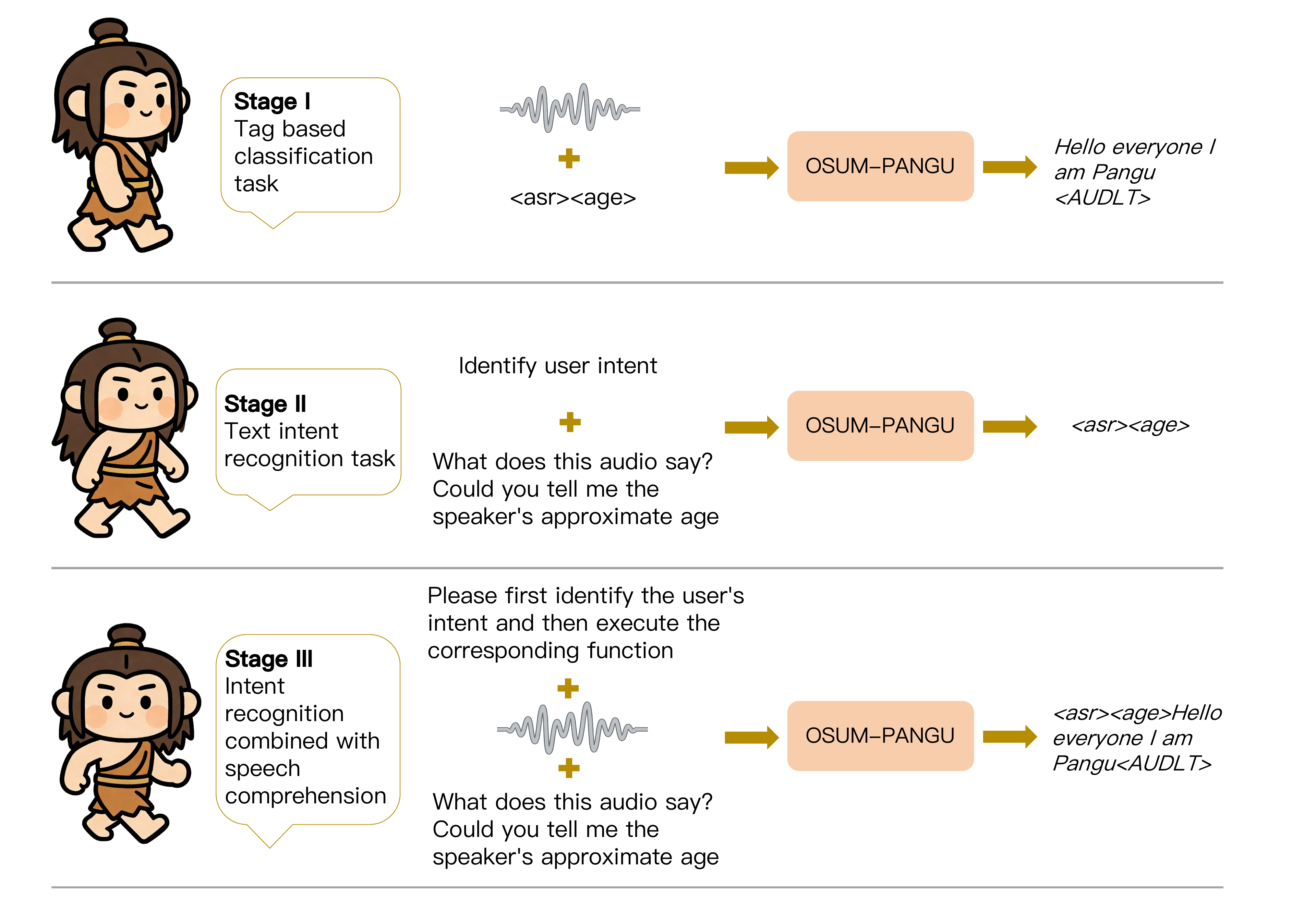}
    }
    \caption{\textbf{The three-stage training pipeline.} 
    Stage I: Tag-based speech alignment. 
    Stage II: Pure text-based intent parsing. 
    Stage III: Joint multimodal integration for intent-driven speech understanding.}
    \label{fig:training_stages}
\end{figure}

\subsection{Training }

To achieve efficient task modality alignment without the need for industrial scale multimodal pre-training, we employ a three-stage training process in Fig.~\ref{fig:training_stages}. This pipeline ensures that the model can interpret spontaneous instructions while maintaining high accuracy in specialized speech tasks.

\textbf{Task Specific Speech Alignment.}
In this stage, the modality adapter is trained while LoRA \cite{lora} is applied to the LLM backbone. Training data includes single-task speech understanding data, multi-task scenarios, and text-based QA pairs to preserve language reasoning ability. Fixed task tags $S_{fixed}$ (e.g., \texttt{<asr>}, \texttt{<age>}) are used as prompts, and the model generates task outputs $Y$ from acoustic tokens $Z_a$:
\begin{equation}
\mathcal{L}_{task} =
-\sum \log P(Y | Z_a, S_{fixed})
\end{equation}
This stage ensures that the adapter extracts task relevant acoustic information that is semantically aligned with the LLM representation space.

\textbf{Text Based Intent Perception.}
This stage trains the LLM to recognize user intent from natural language instructions $S_{nat}$ without audio input. The task is formulated as a text-to-text mapping that converts diverse instruction expressions (e.g., “What does this audio say?”) into structured task identifiers. This stage improves instruction parsing robustness before multimodal integration.

\textbf{Joint Multimodal Integration.}  
In the final stage, the previously learned capabilities are integrated to perform intent-aware speech understanding. The model receives both natural language instructions $S_{nat}$ and raw audio signals $X$, and must autonomously infer the user’s intent and execute the appropriate speech analysis task. The generated output contains both task identifiers and semantic content, such as \texttt{<asr><age>Hello...<ADULT>}. Training minimizes a joint objective:
\begin{equation}
\mathcal{L}_{total} = \mathcal{L}_{intent} + \mathcal{L}_{speech}
\end{equation}
where $\mathcal{L}_{intent}$ supervises the generation of task identifiers to ensure correct intent recognition, and $\mathcal{L}_{speech}$ optimizes the generation of task-specific semantic outputs. This progressive integration allows the LLM to treat natural language instructions as semantic guides for interpreting acoustic tokens $Z_a$.

\subsection{Implementation on NPU Infrastructure}

The entire training and inference pipeline of OSUM-Pangu is implemented on the Ascend NPU platform. Compared with CUDA-based GPU environments commonly used in speech models, this implementation demonstrates the feasibility of deploying multimodal speech understanding systems within a  NPU ecosystem and provides a reference for future non-CUDA research.

\subsection{Instruction Following}

To enable flexible interaction with speech understanding tasks, OSUM-Pangu is designed to interpret and execute user instructions expressed in natural language. Instead of relying on predefined prompts or task tags, the model learns to infer the intended task from user instructions and generate structured outputs. 
To evaluate the model's ability to parse and execute natural language instructions, we introduce the Instruction Following Rate (IFR) as the evaluation metric. IFR measures the proportion of cases in which the model correctly identifies and performs the intended task under diverse natural language prompts.
\begin{equation}
\text{IFR} = \left( \frac{N_{\text{correct}}}{N_{\text{total}}} \right) \times 100\%
\end{equation}
where $N_{\text{correct}}$ denotes the number of instances in which the model output matches the task specified by the user instruction, and $N_{\text{total}}$ represents the total number of evaluation instances.
Since task correctness may involve semantic interpretation beyond simple string matching, we adopt LLM as a Judgefor evaluation. Specifically, DeepSeek-V3 \cite{liu2024deepseek} is used to assess whether the generated output fulfills the user instruction. The judge model evaluates the consistency between the instruction and the response and determines whether the intended task is correctly executed.

\section{Experiments}
\label{sec:experiments}

\subsection{Datasets}
Our experiments follow the task definitions of the OSUM framework~\cite{osum}. To maintain the linguistic reasoning capability of the backbone, we incorporate 2M entries from Alpaca-CoT~\cite{all_t2t} for text-based interactions, with queries synthesized using CosyVoice 2~\cite{tts:cosyvoice2}. To evaluate the model's robustness in real-world scenarios, we utilize an Intent-Instruction Set (IIS) containing over 80k training samples and 4k test prompts, covering diverse colloquial user queries.

\subsection{Setup}

A cornerstone of this work is the non-CUDA software and hardware stack. All experiments were conducted on a cluster of Ascend 910B NPUs using the CANN software stack. The model integrates the openPangu-7B backbone~\cite{pangu} via LoRA~\cite{lora} and a trainable modality adapter. This setup demonstrates the feasibility and efficiency of developing large-scale multimodal speech models within a non-CUDA software hardware ecosystem, providing a reproducible benchmark for the community.

\subsection{Main Results}

The primary goal of our evaluation is to verify whether the localized OSUM-Pangu can achieve performance parity with mainstream models trained on GPU-centric infrastructures. Table \ref{tab:res_multi_public} presents the performance across seven speech-related dimensions.

\begin{table}[ht!]
\centering
\footnotesize
\caption{Evaluation results on public test sets. OSUM-Pangu demonstrates competitive performance across diverse tasks compared to GPU-based baselines Qwen2-Audio and OSUM, proving the effectiveness of the NPU-based pipeline.}
\label{tab:res_multi_public}
\centering
\resizebox{\columnwidth}{!}{%
\begin{tabular}{@{}ccccc@{}}
\toprule
\textbf{Task} & \textbf{Model} & \textbf{Public Test Set} & \textbf{Metric} & \textbf{Public Result} \\ \midrule
\multirow{3}{*}{\textbf{ASR}} & Qwen2-Audio & \multirow{3}{*}{\begin{tabular}[c]{@{}c@{}}WenetSpeech(n/m) \\ AISHELL-2(m/i/a) \\ LibriSpeech (c/o)\end{tabular}} & \multirow{3}{*}{\begin{tabular}[c]{@{}c@{}}WER/CER \\ (\%)\end{tabular}} & \begin{tabular}[c]{@{}c@{}} 8.84 / 8.40 \\ 3.0 / 3.0 / 2.9 \\ \textbf{1.6 / 3.6} \end{tabular} \\ \cmidrule(lr){2-2}
 & OSUM &  &  & \begin{tabular}[c]{@{}c@{}} 6.46 / \textbf{5.34} \\ \textbf{2.81 / 2.75 / 2.73} \\ 2.19 / 5.53 \end{tabular} \\ \cmidrule(lr){2-2}
& \textbf{OSUM-Pangu} &  &  & \begin{tabular}[c]{@{}c@{}} 7.40 / 10.49 \\ 3.01 / 2.98 / 2.95 \\ 3.51 / 8.36 \end{tabular} \\ \midrule

\multirow{3}{*}{\textbf{VED}} & Qwen2-Audio & \multirow{3}{*}{VocalSound} & \multirow{3}{*}{ACC (\%)} & \textbf{93.3} \\ \cmidrule(lr){2-2}
 & OSUM &  &  & 82.58 \\ \cmidrule(lr){2-2}
 & \textbf{OSUM-Pangu} &  &  & 73.04 \\ \midrule

\multirow{3}{*}{\textbf{SER}} & Qwen2-Audio & \multirow{3}{*}{\begin{tabular}[c]{@{}c@{}}MELD-test \\ MER2023\end{tabular}} & \multirow{3}{*}{ACC (\%)} & 55.3 / -- \\ \cmidrule(lr){2-2}
 & OSUM &  &  & 53.38 / 86.43 \\ \cmidrule(lr){2-2}
 & \textbf{OSUM-Pangu} &  &  & 36.40 / \textbf{89.19} \\ \midrule

\multirow{3}{*}{\textbf{SGC}} & Qwen2-Audio & \multirow{3}{*}{Kaggle-CommonVoice test} & \multirow{3}{*}{ACC (\%)} & 97.25 \\ \cmidrule(lr){2-2}
 & OSUM &  &  & \textbf{99.41} \\ \cmidrule(lr){2-2}
 & \textbf{OSUM-Pangu} &  &  & 97.48 \\ \midrule

\multirow{3}{*}{\textbf{SAP}} & Qwen2-Audio & \multirow{3}{*}{Kaggle-CommonVoice test} & \multirow{3}{*}{ACC (\%)} & 35.53 \\ \cmidrule(lr){2-2}
 & OSUM &  &  & 76.52 \\ \cmidrule(lr){2-2}
 & \textbf{OSUM-Pangu} &  &  & \textbf{83.31} \\ \bottomrule
\end{tabular}%
}
\end{table}

As shown in Table \ref{tab:res_multi_public}, OSUM-Pangu achieves task accuracy on par with or even exceeding mainstream GPU-trained models. Notably, in Age Prediction and Style Recognition, our framework exhibits superior performance. While Qwen2-Audio maintains a lead in some ASR benchmarks, the overall results validate that the combination of Ascend NPU and openPangu can support high-fidelity acoustic semantic alignment, bridging the gap between non-CUDA and CUDA AI stacks.

\subsection{Instruction Following Results}

We evaluate the instruction understanding capability of OSUM-Pangu using the IFR. As shown in Table \ref{tab:ifr_results}, OSUM-Pangu achieves an IFR of 90.2\%, outperforming the instruction-tuned baseline Qwen2Audio-Instruct. 
This result indicates that the proposed training strategy effectively enables the model to infer user intent and route tasks correctly under diverse natural language prompts, even without large-scale multimodal instruction pre-training.

\begin{table}[htbp]
\centering
\small
\caption{Comparison of Instruction Following Performance.}
\begin{tabular}{lc}
\toprule
\textbf{Model} & \textbf{IFR (\%)} \\ 
\midrule
Qwen2Audio-Instruct & 71.3  \\
\textbf{OSUM-Pangu (Ours)} & \textbf{90.2}  \\ 
\bottomrule
\end{tabular}
\label{tab:ifr_results}
\end{table}

We further investigate whether natural language instructions introduce performance degradation compared with fixed instruction templates. Table \ref{tab:fi_vs_nl} compares model performance under Fixed Instructions (FI) and Natural Language (NL) prompts across multiple speech understanding tasks.
Results show that the performance difference between FI and NL settings remains minimal for most tasks. In particular, ASR and SER performance remain almost unchanged, indicating that the model can robustly interpret user instructions while maintaining accurate acoustic perception. Although slight degradation is observed in several tasks such as SRWT and VED, the overall performance remains stable, demonstrating that the proposed framework successfully balances instruction flexibility and task accuracy.

\begin{table}[htbp]
\centering
\caption{Performance Comparison: Fixed Instructions (FI) vs. Natural Language (NL).}
\label{tab:fi_vs_nl}
\setlength{\tabcolsep}{2pt} 
\begin{tabular}{cccll}
\toprule
Task & Test & FI & NL & $\Delta$ \\
\midrule
ASR & test-net/librispeech-clean & 7.36/3.64 & 7.40/3.51 & +0.04/-0.13 \\
SER & Test\textsubscript{emotion} & 67.39 & 67.41 & +0.02 \\
SGC & Test\textsubscript{gender} & 97.04 & 96.02 & -1.02 \\
SRWT & Test\textsubscript{align} & 22.39 & 17.52 & -4.87 \\
SSR & Test\textsubscript{style} & 62.79 & 58.05 & -4.74 \\
VED & Test\textsubscript{event} & 77.74 & 73.04 & -4.70 \\
SAP & Test\textsubscript{age} & 71.75 & 72.86 & +0.11 \\
\bottomrule
\end{tabular}
\end{table}

\subsection{Speech-to-Text Chat Results}
The STTC task assesses the reasoning capability of the backbone. As shown in Table \ref{tab:sttc}, while OSUM-Pangu lags behind commercial AI models like ChatGPT-4o, it surpasses several specialized models in TriviaQA and Web Q benchmarks. This highlights the potential of using non-CUDA LLMs for open domain spoken dialogue systems.

\begin{table}[htbp]
\centering
\caption{Comparison on Speech-to-Text Chat (STTC) Benchmarks.}
\label{tab:sttc}
\begin{tabular}{lccc}
\toprule
\textbf{Model} & \textbf{LLaMA Q} & \textbf{TriviaQA} & \textbf{Web Q} \\ \midrule
ChatGPT-4o & 71.7 & 69.7 & 51.6 \\
GLM-4-Voice & 50.7 & 26.5 & 15.9 \\
 DeepTalk & 59.7 & 27.5 &23.1 \\
 OSUM-EChat& 55.3& 33.7&30.4\\
\textbf{OSUM-Pangu} & 44.6 & 28.9 & 29.5 \\ \bottomrule
\end{tabular}
\end{table}

\section{Conclusions}

In this paper, we presented OSUM-Pangu, an open-source  multidimensionspeech understanding framework developed for non-CUDA software hardware ecosystems. By integrating openPangu-7B with the Ascend NPU platform, we demonstrate that alternative AI accelerator infrastructures can achieve performance levels comparable to mainstream GPU-centric frameworks. Through a practical multi-stage training pipeline, OSUM-Pangu effectively bridges the gap between acoustic perception and natural language interaction, achieving a robust instruction following rate of 90.2\% without sacrificing task accuracy. Our work provides a reproducible and high-performance baseline for the open-source community, paving the way for future research on multimodal speech understanding beyond the traditional CUDA ecosystem.

\bibliographystyle{IEEEbib}
\bibliography{main}

@article{tts:cosyvoice2,
  author       = {Zhihao Du and
                  Yuxuan Wang and
                  Qian Chen and
                  Xian Shi and
                  Xiang Lv and
                  Tianyu Zhao and
                  Zhifu Gao and
                  Yexin Yang and
                  Changfeng Gao and
                  Hui Wang and
                  Fan Yu and
                  Huadai Liu and
                  Zhengyan Sheng and
                  Yue Gu and
                  Chong Deng and
                  Wen Wang and
                  Shiliang Zhang and
                  Zhijie Yan and
                  Jingren Zhou},
  title        = {CosyVoice 2: Scalable Streaming Speech Synthesis with Large Language
                  Models},
  journal      = {CoRR},
  volume       = {abs/2412.10117},
  year         = {2024}
}

@article{um:qwen-audio,
  author       = {Yunfei Chu and
                  Jin Xu and
                  Xiaohuan Zhou and
                  Qian Yang and
                  Shiliang Zhang and
                  Zhijie Yan and
                  Chang Zhou and
                  Jingren Zhou},
  title        = {Qwen-Audio: Advancing Universal Audio Understanding via Unified Large-Scale
                  Audio-Language Models},
  journal      = {CoRR},
  volume       = {abs/2311.07919},
  year         = {2023}
}

@article{um:qwen2audio,
  author       = {Yunfei Chu and
                  Jin Xu and
                  Qian Yang and
                  Haojie Wei and
                  Xipin Wei and
                  Zhifang Guo and
                  others},
  title        = {Qwen2-Audio Technical Report},
  journal      = {CoRR},
  volume       = {abs/2407.10759},
  year         = {2024}
}

@article{llm:llama3,
  author       = {Abhimanyu Dubey and Abhinav Jauhri and Abhinav Pandey and
                  Abhishek Kadian and others},
  title        = {The Llama 3 Herd of Models},
  journal      = {arXiv preprint arXiv:2407.21783},
  year         = {2024},
  url          = {https://arxiv.org/abs/2407.21783}
}

@article{llm:gpt4,
  author       = {OpenAI},
  title        = {{GPT-4} Technical Report},
  journal      = {CoRR},
  volume       = {abs/2303.08774},
  year         = {2023}
}

@inproceedings{asr:whisper,
  author       = {Alec Radford and
                  Jong Wook Kim and
                  Tao Xu and
                  others},
  title        = {Robust Speech Recognition via Large-Scale Weak Supervision},
  booktitle    = {ICML},
  volume       = {202},
  pages        = {28492--28518},
  publisher    = {{PMLR}},
  year         = {2023}
}

@inproceedings{transformer,
  author       = {Ashish Vaswani and
                  Noam Shazeer and
                  Niki Parmar and
                  Jakob Uszkoreit and
                  Llion Jones and
                  Aidan N. Gomez and
                  Lukasz Kaiser and
                  Illia Polosukhin},
  editor       = {Isabelle Guyon and
                  Ulrike von Luxburg and
                  Samy Bengio and
                  Hanna M. Wallach and
                  Rob Fergus and
                  S. V. N. Vishwanathan and
                  Roman Garnett},
  title        = {Attention is All you Need},
  booktitle    = {Advances in Neural Information Processing Systems 30: Annual Conference
                  on Neural Information Processing Systems 2017, December 4-9, 2017,
                  Long Beach, CA, {USA}},
  pages        = {5998--6008},
  year         = {2017}
}

@article{osum,
  author       = {Xuelong Geng and
                  Kun Wei and
                  Qijie Shao and
                  Shuiyun Liu and
                  Zhennan Lin and
                  Zhixian Zhao and
                  Guojian Li and
                  Wenjie Tian and
                  Peikun Chen and
                  Yangze Li and
                  Pengcheng Guo and
                  Mingchen Shao and
                  Shuiyuan Wang and
                  Yuang Cao and
                  Chengyou Wang and
                  Tianyi Xu and
                  Yuhang Dai and
                  Xinfa Zhu and
                  Yue Li and
                  Li Zhang and
                  Lei Xie},
  title        = {{OSUM:} Advancing Open Speech Understanding Models with Limited Resources
                  in Academia},
  journal      = {CoRR},
  volume       = {abs/2501.13306},
  year         = {2025}
}

@article{osum-echat,
  author       = {Xuelong Geng and
                  Qijie Shao and
                  Hongfei Xue and
                  Shuiyuan Wang and
                  Hanke Xie and
                  Zhao Guo and
                  Yi Zhao and
                  Guojian Li and
                  Wenjie Tian and
                  Chengyou Wang and
                  Zhixian Zhao and
                  Kangxiang Xia and
                  Ziyu Zhang and
                  Zhennan Lin and
                  Tianlun Zuo and
                  Mingchen Shao and
                  Yuang Cao and
                  Guobin Ma and
                  Longhao Li and
                  Yuhang Dai and
                  Dehui Gao and
                  Dake Guo and
                  Lei Xie},
  title        = {OSUM-EChat: Enhancing End-to-End Empathetic Spoken Chatbot via Understanding-Driven Spoken Dialogue},
  journal      = {CoRR},
  volume       = {abs/2508.09600},
  year         = {2025}
}

@inproceedings{all_t2t,
  author       = {Qingyi Si and
                  Tong Wang and
                  Zheng Lin and
                  Xu Zhang and
                  others},
  title        = {An Empirical Study of Instruction-tuning Large Language Models in
                  Chinese},
  booktitle    = {EMNLP},
  pages        = {4086--4107},
  year         = {2023}
}

@article{CNN,
  author       = {Keiron O'Shea and
                  Ryan Nash},
  title        = {An Introduction to Convolutional Neural Networks},
  journal      = {CoRR},
  volume       = {abs/1511.08458},
  year         = {2015}
}

@article{audio_flamingo3,
  author       = {Arushi Goel and
                  Sreyan Ghosh and
                  Jaehyeon Kim and
                  Sonal Kumar and
                  Zhifeng Kong and
                  others},
  title        = {Audio Flamingo 3: Advancing Audio Intelligence with Fully Open Large
                  Audio Language Models},
  journal      = {CoRR},
  volume       = {abs/2507.08128},
  year         = {2025},
}

@article{pangu,
  title={Pangu embedded: An efficient dual-system llm reasoner with metacognition},
  author={Chen, Hanting and Wang, Yasheng and Han, Kai and Li, Dong and Li, Lin and Bi, Zhenni and others},
  journal={arXiv preprint arXiv:2505.22375},
  year={2025}
}

@inproceedings{salmonn,
  author       = {Changli Tang and
                  Wenyi Yu and
                  Guangzhi Sun and
                  Xianzhao Chen and
                  Tian Tan and
                  others},
  title        = {{SALMONN:} Towards Generic Hearing Abilities for Large Language Models},
  booktitle    = {ICLR},
  year         = {2024},
}

@article{simple-asr-llm,
  author       = {Ziyang Ma and
                  Guanrou Yang and
                  Yifan Yang and
                  Zhifu Gao and
                  Jiaming Wang and
                  others},
  title        = {{An Embarrassingly Simple Approach for {LLM} with Strong {ASR} Capacity}},
  journal      = {{CoRR}},
  year         = {2024},
}

@inproceedings{lora,
  author       = {Edward J. Hu and
                  Yelong Shen and
                  others},
  title        = {{LoRA: Low-Rank Adaptation of Large Language Models}},
  booktitle    = {{ICLR}
                  },
  year         = {2022},
}

@inproceedings{asr_my,
  author       = {Xuelong Geng and
                  Tianyi Xu and
                  Kun Wei and
                  Bingshen Mu and
                  Hongfei Xue and
                  He Wang and
                  Yangze Li and
                  Pengcheng Guo and
                  Yuhang Dai and
                  Longhao Li and
                  Mingchen Shao and
                  Lei Xie},
  title        = {Unveiling the Potential of LLM-Based {ASR} on Chinese Open-Source
                  Datasets},
  booktitle    = {ISCSLP},
  pages        = {26--30},
  publisher    = {{IEEE}},
  year         = {2024},
  url          = {https://doi.org/10.1109/ISCSLP63861.2024.10800077},
  doi          = {10.1109/ISCSLP63861.2024.10800077},
  bibsource    = {dblp computer science bibliography, https://dblp.org}
}

@article{gong2023ltu,
  title={Listen, think, and understand},
  author={Gong, Yuan and Luo, Hongyin and others},
  journal={arXiv preprint arXiv:2305.10790},
  year={2023}
}

@inproceedings{gong2023ltu-as,
  title={Joint audio and speech understanding},
  author={Gong, Yuan and Liu, Alexander H and Luo, Hongyin and others},
  booktitle={ASRU},
  pages={1--8},
  year={2023},
  organization={IEEE}
}

@article{deshmukh2023pengi,
  title={Pengi: An audio language model for audio tasks},
  author={Deshmukh, Soham and Elizalde, Benjamin and Singh, Rita and Wang, Huaming},
  journal={Advances in Neural Information Processing Systems},
  volume={36},
  pages={18090--18108},
  year={2023}
}

@inproceedings{ghosh2024gama,
  title={Gama: A large audio-language model with advanced audio understanding and complex reasoning abilities},
  author={Ghosh, Sreyan and Kumar, Sonal and Seth, Ashish and others},
  booktitle={EMNLP},
  pages={6288--6313},
  year={2024}
}

@article{abouelenin2025phi,
  title={Phi-4-mini technical report: Compact yet powerful multimodal language models via mixture-of-loras},
  author={Abouelenin, Abdelrahman and Ashfaq, Atabak and Atkinson, Adam and Awadalla, Hany and Bach, Nguyen and others},
  journal={arXiv preprint arXiv:2503.01743},
  year={2025}
}

@article{team2023gemini,
  title={Gemini: a family of highly capable multimodal models},
  author={Team, Gemini and Anil, Rohan and Borgeaud, Sebastian and Alayrac, Jean-Baptiste and Yu, Jiahui and others},
  journal={arXiv preprint arXiv:2312.11805},
  year={2023}
}

@article{hurst2024gpt,
  title={Gpt-4o system card},
  author={Hurst, Aaron and Lerer, Adam and Goucher, Adam P and others},
  journal={arXiv preprint arXiv:2410.21276},
  year={2024}
}

@article{kong2024audioflamingo,
  title={Audio flamingo: A novel audio language model with few-shot learning and dialogue abilities},
  author={Kong, Zhifeng and Goel, Arushi and others},
  journal={arXiv preprint arXiv:2402.01831},
  year={2024}
}

@article{kong2025audioflamingo2,
  title={Audio flamingo 2: An audio-language model with long-audio understanding and expert reasoning abilities},
  author={Ghosh, Sreyan and Kong, Zhifeng and Kumar, Sonal and others},
  journal={arXiv preprint arXiv:2503.03983},
  year={2025}
}

@article{yin2025pangu,
  title={Pangu ultra: Pushing the limits of dense large language models on ascend npus},
  author={Yin, Yichun and Huang, Wenyong and Song, Kaikai and Tang, Yehui and Wu, Xueyu and others},
  journal={arXiv preprint arXiv:2504.07866},
  year={2025}
}

@inproceedings{zhu2023performance,
  title={Performance evaluation of mindspore and pytorch based on ascend npu},
  author={Zhu, Zeling and Wang, Bangchuan and Yang, Chuying and Zhu, Rui and Zhou, Mingyao and Zheng, Nenggan},
  booktitle={ICPADS},
  pages={1826--1832},
  year={2023},
  organization={IEEE}
}

@inproceedings{wang2023analysis,
  title={Analysis of performance and optimization in mindspore on ascend npus},
  author={Wang, Bangchuan and Yang, Chuying and Zhu, Rui and Liu, Xiao and Zhou, Mingyao and Zheng, Nenggan},
  booktitle={ICPADS},
  pages={1701--1708},
  year={2023},
  organization={IEEE}
}

@article{yang2024qwen2technicalreport,
  author       = {An Yang and
                  Baosong Yang and
                  Binyuan Hui and
                  Bo Zheng and
                  others},
  title        = {Qwen2 Technical Report},
  journal      = {CoRR},
  volume       = {abs/2407.10671},
  year         = {2024},
}

@article{liu2024deepseek,
  title={Deepseek-v3 technical report},
  author={Liu, Aixin and Feng, Bei and Xue, Bing and Wang, Bingxuan and Wu, Bochao and Lu, Chengda and Zhao, Chenggang and Deng, Chengqi and Zhang, Chenyu and Ruan, Chong and others},
  journal={arXiv preprint arXiv:2412.19437},
  year={2024}
}

@inproceedings{xue2025enhancing,
  title={Enhancing non-core language instruction-following in speech llms via semi-implicit cross-lingual cot reasoning},
  author={Xue, Hongfei and Tang, Yufeng and Liu, Hexin and Zhang, Jun and Geng, Xuelong and Xie, Lei},
  booktitle={ACMMM},
  pages={10984--10993},
  year={2025}
}

\end{document}